\documentstyle{mn}
\input{epsf.tex}
\newif\ifAMStwofonts

\ifoldfss
 \ifCUPmtlplainloaded \else
  \NewTextAlphabet{textbfit} {cmbxti10} {}
  \NewTextAlphabet{textbfss} {cmssbx10} {}
  \NewMathAlphabet{mathbfit} {cmbxti10} {} 
  \NewMathAlphabet{mathbfss} {cmssbx10} {} 
 \fi %
 \ifAMStwofonts
  \ifCUPmtlplainloaded \else
   \NewSymbolFont{upmath} {eurm10}
   \NewSymbolFont{AMSa} {msam10}
   \NewMathSymbol{\upi}   {0}{upmath}{19}
   \NewMathSymbol{\umu}   {0}{upmath}{16}
   \NewMathSymbol{\upartial}{0}{upmath}{40}
   \NewMathSymbol{\leqslant}{3}{AMSa}{36}
   \NewMathSymbol{\geqslant}{3}{AMSa}{3E}

    \let\le=\leqslant
    \let\ge=\geqslant
  \fi
 \fi
\fi 

\ifnfssone
 \newmathalphabet{\mathit}
 \addtoversion{normal}{\mathit}{cmr}{m}{it}
 \addtoversion{bold}{\mathit}{cmr}{bx}{it}
 \newmathalphabet{\mathbfit} 
 \addtoversion{normal}{\mathbfit}{cmr}{bx}{it}
 \addtoversion{bold}{\mathbfit}{cmr}{bx}{it}
 \newmathalphabet{\mathbfss} 
 \addtoversion{normal}{\mathbfss}{cmss}{bx}{n}
 \addtoversion{bold}{\mathbfss}{cmss}{bx}{n}
 \ifAMStwofonts
  \ifCUPmtlplainloaded \else
   \UseAMStwoboldmath
   \makeatletter
   \new@mathgroup\upmath@group
   \define@mathgroup\mv@normal\upmath@group{eur}{m}{n}
   \define@mathgroup\mv@bold\upmath@group{eur}{b}{n}
   \edef\UPM{\hexnumber\upmath@group}
   \new@mathgroup\amsa@group
   \define@mathgroup\mv@normal\amsa@group{msa}{m}{n}
   \define@mathgroup\mv@bold\amsa@group{msa}{m}{n}
   \edef\AMSa{\hexnumber\amsa@group}
   \makeatother
   \mathchardef\upi="0\UPM19
   \mathchardef\umu="0\UPM16
   \mathchardef\upartial="0\UPM40
   \mathchardef\leqslant="3\AMSa36
   \mathchardef\geqslant="3\AMSa3E

    \let\le=\leqslant
    \let\ge=\geqslant
  \fi
 \fi
\fi 

\ifnfsstwo
 \DeclareMathAlphabet{\mathbfit}{OT1}{cmr}{bx}{it}
 \SetMathAlphabet\mathbfit{bold}{OT1}{cmr}{bx}{it}
 \DeclareMathAlphabet{\mathbfss}{OT1}{cmss}{bx}{n}
 \SetMathAlphabet\mathbfss{bold}{OT1}{cmss}{bx}{n}
 \ifAMStwofonts
  \ifCUPmtlplainloaded \else
   \DeclareSymbolFont{UPM}{U}{eur}{m}{n}
   \SetSymbolFont{UPM}{bold}{U}{eur}{b}{n}
   \DeclareSymbolFont{AMSa}{U}{msa}{m}{n}
   \DeclareMathSymbol{\upi}{0}{UPM}{"19}
   \DeclareMathSymbol{\umu}{0}{UPM}{"16}
   \DeclareMathSymbol{\upartial}{0}{UPM}{"40}
   \DeclareMathSymbol{\leqslant}{3}{AMSa}{"36}
   \DeclareMathSymbol{\geqslant}{3}{AMSa}{"3E}

    \let\le=\leqslant
    \let\ge=\geqslant
  \fi
 \fi
\fi 

\ifCUPmtlplainloaded \else
 \ifAMStwofonts \else 
  \def\upi{\pi}
  \def\umu{\mu}
  \def\upartial{\partial}
 \fi
\fi

\title{Orbital structure in oscillating galactic potentials}

\author[B. Terzi{\'c} and H. E. Kandrup]
 {Bal{\v s}a Terzi{\'c}$^{1}$\thanks{E-mail: bterzic@astro.ufl.edu} and
 Henry E. Kandrup$^{1,2,3}$\thanks{E-mail: kandrup@astro.ufl.edu}\\
  $^{1}$ Department of Astronomy, University of Florida, Gainesville, 
     FL 32611-2055, USA\\
  $^{2}$ Department of Physics, University of Florida, Gainesville, 
     FL 32611-2055, USA\\
  $^{3}$ Institute for Fundamental Theory, University of Florida, Gainesville,
     FL 32611-2055, USA}

\date{Accepted 2003 October 8.
   Received 2003 October 8; in the original form 2003 June 22 }

\pagerange{\pageref{firstpage}--\pageref{lastpage}}
\pubyear{2003}

\begin{document}

\maketitle

\label{firstpage}

\begin{abstract}
Subjecting a galactic potential to (possibly damped) nearly periodic, 
time-dependent variations can lead to large numbers of chaotic orbits 
experiencing systematic changes in energy, and the resulting chaotic phase 
mixing could play an important role in explaining such phenomena as violent 
relaxation. This paper focuses on the simplest case of spherically symmetric 
potentials subjected to strictly periodic driving with the aim of understanding
precisely why orbits become chaotic and under what circumstances they will 
exhibit systematic changes in energy. Four unperturbed potentials 
$V_{0}({\bf r})$ were considered, each subjected to a time-dependence of the form
$V({\bf r},t)=V_{0}({\bf r})(1+m_{0}\sin {\omega}t)$. 
In each case, the orbits divide clearly into regular and chaotic, distinctions 
which appear absolute. In particular, transitions from regularity to chaos 
are seemingly impossible.
Over finite time intervals, chaotic orbits subdivide into what can be termed
`sticky' chaotic orbits, which exhibit no large scale secular changes in
energy and remain trapped in the phase space region where they started; and
`wildly' chaotic orbits, which {\em do} exhibit systematic drifts in energy 
as the orbits diffuse to different phase space regions. This latter
distinction is not absolute, transitions corresponding apparently to orbits 
penetrating a `leaky' phase space barrier. 
The three different orbit types can be identified simply in terms of the
frequencies for which their Fourier spectra have the most power.
An examination of the statistical properties of orbit ensembles as a function
of driving frequency ${\omega}$ allows one to identify the specific resonances 
that determine orbital structure. 
Attention focuses also on how, for fixed amplitude $m_{0}$, such quantities
as the mean energy shift, the relative measure of chaotic orbits, and the
mean value of the largest Lyapunov exponent vary with driving frequency
${\omega}$; and how, for fixed ${\omega}$, the same quantities depend on
$m_{0}$.
\end{abstract}


\section{INTRODUCTION AND MOTIVATION}
The introduction of a periodic, or nearly periodic, time-dependence into even
an integrable potential can lead to large measures of chaotic orbits and,
in many cases, significant systematic changes in 
energy (Gluckstern 1994, Kandrup, Vass \& Sideris 2003, Kandrup, Sideris,
Terzi{\'c} \& Bohn 2003); and the resulting chaotic phase mixing can drive 
an efficient shuffling of orbits in phase space. 
This effect, when associated with the self-consistent bulk potential 
arising in a many-body system interacting via long range Coulomb forces,
could help explain diverse phenomena including violent relaxation (Lynden-Bell
1967) in galaxies and various collective effects observed in charged particle
beams (Kandrup 2003).

The objective here is to study this effect at the level of individual orbits
with the aim of determining as precisely as possible the origins of this
efficient phase mixing. Why is it that a nearly periodic time-dependence 
can make many, albeit not necessarily all, orbits chaotic? And under what
circumstances will the orbits experience large changes in energy?

A seemingly paradigmatic example is provided by a harmonic
oscillator potential with a time-dependent frequency ${\Omega}(t)$,
\begin{equation}
{d^{2}x\over dt^{2}}+{\Omega}^{2}(t)x=0, 
\qquad {\rm with} \qquad
{\Omega}^{2}=a+b\cos{\omega}t, 
\end{equation}
which can be rescaled into a standard Mathieu equation (see, e.g., Matthews
\& Walker 1970)
\begin{equation}
{d^{2}x\over d{\tau}^{2}}+\left({\alpha}+{\beta}\cos 2{\tau}\right)x = 0.
\end{equation}
This example suggests (i) that this time-dependent chaos is resonant in 
origin; (ii) that its strongest manifestations should be associated with a 
$2:1$ resonance between the natural and driving frequencies; and (iii) that 
increasing the amplitude of the time-dependence should increase the range 
in driving frequencies that yield chaotic orbits as well as the relative 
measure of chaotic orbits and the size of a typical Lyapunov exponent at any 
given frequency. 

Superficially, such predictions would seem correct. If, for example, the power
spectrum generated from a time series of the acceleration experienced by 
unperturbed orbits has power concentrated primarily in frequencies satisfying
${\Omega}_{1}{\;}{\la}{\;}{\Omega}{\;}{\la}{\;}{\Omega}_{2}$, the frequencies
${\omega}$ for which an imposed driving will elicit a significant response
often satisfy, at least approximately, 
${\Omega}_{1}{\;}{\la}{\;}{\omega}/2{\;}{\la}{\;}{\Omega}_{2}$
({\em e.g.,} Kandrup, Vass, and Sideris 2003). 
Moreover, one finds that, as probed by both the relative measure of strongly
chaotic orbits and the size of a typical Lyapunov exponent, the amount and
degree of chaos are increasing functions of perturbation amplitude. For
sufficiently low amplitudes, there can be a comparatively sensitive dependence
on driving frequency. As amplitude is increased, however, the resonances 
broaden and much of this `fine structure' is lost.

This is illustrated in Fig.~1, which was generated for ensembles
of $N=800$ orbits with energies $E=-0.5$, evolved for a fixed time $t=256$ in 
Plummer and ${\gamma}=1$ Dehnen (1993) potentials that were subjected to a 
time-dependent periodic driving 
{\em i.e.,} $V({\bf r},t)=V_{0}({\bf r})(1+m_{0}\sin {\omega}t)$
with variable frequencies ${\omega}$.
The top two panels exhibit ${\langle}{\chi}{\rangle}$, the mean value of the 
largest finite time Lyapunov exponent for the ensemble.
The bottom two panels exhibit the `average' power 
spectra $|a_{x}({\Omega})|$ for the the $x$-component of the acceleration
experienced by the orbits when evolved in the absence of any driving (the
$y$- and $z$-components are statistically identical):
\begin{equation}
|a_{x}({\Omega})|^{2}=\sum_{i=1}^{N}|a_{x,i}({\Omega})|^{2}.
\end{equation}

This, however, is not the whole story! Detailed investigations of the 
dependence on driving frequency indicate that, even though the overall
response seems especially strong {\em near} various resonances, for example
for ${\omega}{\;}{\sim}{\;}{\Omega}_{0}/2$, 
with ${\Omega}_{0}$ the peak frequency associated with $|a_{x}({\Omega})|$
and $|x({\Omega})|$, the resonant frequency can actually constitute a 
(near-){\em local minimum} in the amount and degree of chaos. Pulsing
orbits with that precise frequency can actually render them especially stable!

The oscillator example might also suggest that chaotic orbits typically
exhibit exponentially fast changes in energy. This, however, is not the case.
In general, chaos does {\em not} trigger exponentially fast systematic changes
in energy, at least as viewed macroscopically. Shuffling of 
orbits in terms of energy is less efficient overall than shuffling in such
orbital elements as position or velocity. This is consistent with 
numerical simulations of galaxy formation and galaxy-galaxy collisions which 
indicate that, even during comparatively `violent' processes, stars can 
exhibit a significant remembrance of their initial binding energies (see, 
{\em e.g.}, Quinn \& Zurek 1988, Kandrup, Mahon \& Smith 1993).

The specific aim of this paper is to focus on ensembles of orbits with fixed
initial energy evolved in potentials which, in the absence of any 
perturbations, are spherically symmetric and hence integrable; and to explore
as a function of driving frequency the effects of a spherically symmetric 
time-dependent perturbation in terms of the power spectra, finite time
Lyapunov exponents, and energy shifts of individual orbits. 

The results are based on an examination of orbits evolved in four different 
potentials subjected to strictly periodic, spherically symmetric oscillations. 
Section II focuses on the origins of chaos in these potentials, demonstrating 
in particular that the orbits divide naturally into three populations, 
corresponding to (i) regular orbits; (ii) `sticky' chaotic orbits which exhibit 
little if any systematic changes in energy and remain confined to a relatively 
small phase space region; and (iii) `wildly' chaotic orbits which can exhibit 
large systematic changes in energy as they diffuse to different phase space 
regions. The distinction between regular and chaotic appears enforced by 
absolute barriers.  By contrast, the distinction between `sticky' and `wildly' 
chaotic orbits is only approximate, seemingly enforced by entropy barriers 
which appear to behave in much the same fashion as cantori in time-independent 
two-degree-of-freedom Hamiltonian systems.

Section III uses a frequency space analysis to demonstrate that orbits in
a time-periodic potential, both regular and chaotic, arise from
resonances involving the driving frequency ${\omega}$. In particular, 
an investigation of Fourier spectra as a function of driving frequency 
demonstrates that, as ${\omega}$ is increased from zero, there is a 
progression through successively higher order resonances.

Most of the work in Sections II and III focuses on a single, comparatively low
amplitude -- $m_{0}=0.2$ -- time-dependence, sufficiently small that resonances 
are comparatively narrow and, hence, relatively easily interpreted. Section IV
extends the analysis by considering how such bulk properties as the relative
abundances of regular, `sticky', and `wildly' chaotic orbits depend on the 
amplitude $m_{0}$ of the perturbation.  Section V concludes by speculating 
on implications for violent relaxation and other collective processes in real 
galaxies.  

Finally, a brief glossary may prove useful. In what follows, ${\omega}$ refers
to driving frequency, ${\Omega}$ to the frequency associated with
a Fourier transform. ${\Omega}_{0}(E)$ denotes the peak frequency, {\em  i.e.,} 
the frequency containing the most power, in the Fourier transform of some orbital 
quantity as evaluated for an ensemble of initial conditions with energy $E$ 
evolved with ${\omega}=0$. ${\Omega}^{(1)}$ and ${\Omega}^{(2)}$ denote, 
respectively, the frequencies containing the largest and second largest amounts 
of power for a single initial condition evolved with a (in general) nonzero 
${\omega}$.  ${\Omega}^{(1)}_{0}$ and ${\Omega}^{(2)}_{0}$ denote the same 
quantities for the same initial condition evolved with ${\omega}=0$.

\section{REGULAR AND CHAOTIC ORBITS}

The results derived in this paper derive from an analysis of four 
potentials, namely a pulsed Plummer potential, 
\begin{equation}
V(r,t)=-{m(t)\over (1+r^{2})^{1/2}},
\end{equation}
and pulsed Dehnen potentials, 
\begin{equation}
V(r,t)=-{m(t)\over (2-{\gamma})}
\left[1- {r^{2-{\gamma}}\over (1+r)^{2-{\gamma}}} \right]
\end{equation}
for ${\gamma}=0.0$, $0.5$, and $1.0$. In each case,
$r^{2}=x^{2}+y^{2}+z^{2}$ and 
\begin{equation}
m(t)=1+m_{0}\sin{\omega}t.
\end{equation}
Both the Plummer and the ${\gamma}=0$ Dehnen potentials reduce to a harmonic 
oscillator for $r\to 0$, this corresponding to a constant density core. 
The ${\gamma}=0.5$ and
${\gamma}=1.0$ Dehnen potentials yield cuspy density profiles. For the Plummer
potential the lowest possible unpulsed energy $E_{min}=-1.0$. For the Dehnen
potential, $E_{min}$ depends on ${\gamma}$.

The analysis involved selecting representative ensembles of $N{\;}{\ge}{\;}800$
initial conditions with fixed initial energy $E$ and exploring the properties 
of orbits generated by evolving these with different driving frequencies 
${\omega}$. In most cases `representative' was interpreted as corresponding 
to a microcanonical distribution, {\em i.e.,} a uniform sampling of the 
constant energy hypersurface, generated using an
algorithm described in Kandrup \& Siopis (2003). What this entailed was:
(i) sampling the energetically accessible configuration space regions, for
which the unperturbed potential $V({\bf r}){\;}{\le}{\;}E$, with weight
proportional to $(E-V)^{1/2}$ (in general, for a $D$-dimensional potential,
the weight would be $(E-V)^{(D-2)/2}$); and then (ii) assigning to each 
selected phase space point a randomly oriented velocity with magnitude 
$|{\bf v}|=[2(E-V)]^{1/2}$.

The first obvious objective is to quantify the total amount of chaos at
different initial energies $E$ as a function of driving frequency ${\omega}$.
This was done by integrating the equations of motion for each initial 
condition for a fixed interval $t=512$ while simultaneously computing an 
estimate of the largest (finite time) Lyapunov exponent, and then extracting 
the mean value ${\langle}{\chi}{\rangle}$ for all the orbits. 

A plot of ${\langle}{\chi}{\rangle}$ as a function of ${\omega}$ exhibits
considerable structure. For sufficiently low and high frequencies, the driving
has only a minimal effect: the orbits remain regular with the computed
${\langle}{\chi}{\rangle}$ very small. For intermediate frequencies, however, 
comparable to the natural frequencies for which the acceleration 
has appreciable power, ${\langle}{\chi}{\rangle}$ will increase from its
near-zero value, this indicating that a significant fraction of the initial
conditions correspond to chaotic orbits with exponentially sensitive dependence
on initial conditions. Also evident is the fact that the functional dependence 
of ${\langle}{\chi}{\rangle}$ on ${\omega}$ can be quite complex. 
Fig.~1 (a) and (b) provides simple examples of this behaviour.

That the locations of the peaks and troughs in ${\langle}{\chi}{\rangle}$  
are related to the natural frequencies of the unperturbed orbits can be seen 
if, in a plot of ${\langle}{\chi}{\rangle}({\omega})$, is superimposed the 
Fourier spectrum of the radial acceleration $|a_{r}({\Omega})|$ and/or the 
$x$-component $|a_{x}({\Omega})|$ for the unperturbed ensemble. 
Examples thereof are provided in Fig.~2, which focuses in greater detail on 
lower frequencies ${\le}{\;}6.0$.  Here $|a_{r}|$ is defined by analogy with 
eq.~(3) except that the large zero frequency contribution reflecting a nonzero 
average $r$ has been subtracted.
(The bottom row of this Figure is discussed in Section III.)
There are direct correlations between $|a_{x}|$ (or $|a_{r}|$) and  
${\langle}{\chi}{\rangle}$ in the sense, for example, that the spacing between 
successive peaks for these two quantities are often comparable in magnitude. 
However, it is not true that (say) peaks in $|a_{x}|$ coincide with peaks in 
${\langle}{\chi}{\rangle}$. Something more complex must be going on. 

To address these complications and, especially, to identify which orbits will
be chaotic, it proves convenient to classify the orbits in terms of properties 
of their Fourier spectra. Here the most obvious tack is to determine the value 
of the `peak' frequency ${\Omega}^{(1)}$ for each orbit, {\em i.e.,} the 
value of ${\Omega}$ for which its $|x({\Omega})|$ (and, by symmetry, 
$|y({\Omega})|$ and  $|z({\Omega})|$) is maximised, and to look 
for correlations between ${\Omega}^{(1)}$ and the finite time Lyapunov 
exponent ${\chi}$ for the orbit. 

Far from the resonant regions, at both higher and lower frequencies, all the 
orbits are regular, with small values of ${\chi}$ and peak frequencies 
${\Omega}^{(1)}$ relatively close to the peak frequency ${\Omega}^{(1)}_{0}$
associated with an ${\omega}=0$ orbit with the same initial condition. 
Closer, however, to the resonance one sees the onset of
chaos, in which some of these regular orbits undergo a significant
change in the form of their spectra. 

Consider, for example the observed behaviour of orbits in a given ensemble
as the driving frequency ${\omega}$ is slowly increased from zero to higher
frequencies. The introduction of a small but nonzero ${\omega}$ alters slightly
the peak frequency in the Fourier spectra $|x({\Omega})|$,
but all the orbits still remain regular. Nevertheless, the spectra {\em do} 
become more complicated, typically acquiring additional power at one or more 
lower frequencies. 

For sufficiently small ${\omega}$ no significant resonant couplings are
triggered.  However, there is an abrupt onset of chaos when 
${\omega}$ becomes sufficiently large that, for some of the orbits, one 
or more of these new lower frequencies acquires more power than what had been 
the higher peak frequency ${\Omega}^{(1)}$. At this stage, resonant couplings 
become important and the orbits divide generically into three different 
populations:
Population I regular orbits, which behave much the same as they did at lower
driving frequencies ${\omega}$; Population II chaotic orbits, where the peak 
frequency ${\Omega}^{(1)}$ in the Fourier spectrum is close in value to half 
the driving frequency; and Population III chaotic orbits, where 
${\Omega}^{(1)}$ is substantially lower and tends to drift in time. 

Examples of this behaviour for the Plummer and ${\gamma}=0$
Dehnen potentials are provided in Fig.~3, which exhibit scatter plots of 
finite time Lyapunov exponent and peak frequency ${\Omega}^{(1)}$ at 
four times, $t=1024$, $2048$, $4096$, and $8192$.

The Population I orbits are clearly regular. Plots of the Fourier spectra
for spatial coordinates and velocities
exhibit sharply defined peaks and the computation of a finite time Lyapunov
exponent  ${\chi}(t)$ results in a uniform decay of the form expected
(see, e.g., Bennetin, Galgani \& Strelcyn 1976) for regular orbits. It is,
for example, evident from Fig.~3 that the typical ${\chi}$ for orbits with
${\Omega}^{(1)}>{\omega}/2$ decreases as time passes. Equally clearly,
the Population II and III orbits are chaotic: Their Fourier
spectra are broader band, and the finite time Lyapunov exponent ${\chi}(t)$'s, 
typically substantially larger in magnitude, do not decay towards zero
systematically. 

When ${\omega}$ is nonzero, energy is (almost) never conserved since the
potential is time-dependent. However, the form of the variations in energy
is different for the three different orbit populations. Regular
Population I orbits show completely periodic oscillations in energy, the 
Fourier spectra of $E(t)$ being excedingly simple and regular. For chaotic 
Population II orbits, changes in energy are again bounded and are again
nearly oscillatory, although occasional `glitches' are observed that break
near-periodicity. The Fourier spectra are more complex than for the regular 
orbits. However, they typically assume forms which one is wont to associate 
with nonconserved phase space coordinates for a `sticky' near-regular orbit 
in a time-independent potential. For the Population III chaotic orbits, the 
behaviour is very different. In this case, the time-dependence of the energy
is far from periodic and the spectrum looks manifestly chaotic. For these
orbits, the energy typically exhibits large secular variations, drifting to
significantly larger or smaller values. 

The distinction between regular and chaotic orbits appears to be absolute. If 
an initial condition corresponding to a regular orbit is integrated even for
a very long time, the orbit will remain regular, with small ${\chi}$, sharply
peaked Fourier spectra, and phase space coordinates confined to the same
small region. By contrast, the distinction between the two types of chaotic
orbits is {\em not} absolute. It appears that, if one integrates long enough 
-- in some cases a time $t{\;}{\gg}{\;}100000$ may be required! -- many 
(if not all) Population II orbits originally trapped near the initial regular 
region escape to become Population III orbits. 
The possibility of such transitions suggests strongly that the boundary
between Populations II and III corresponds to an entropy barrier, {\em i.e.,}
a leaky phase space barrier through which escape is possible but which 
requires an orbit to `find a small gap'.

Examples of the aforementioned behaviour are illustrated in Fig.~4, which
was constructed for representative orbits in a pulsed Plummer potential.
Here the left hand columns exhibits the phase space coordinates $x$ and 
$v_{x}$ of four representative orbits,
each recorded at intervals ${\delta}t=1.0$. 
The second column exhibits the mean shift in energy, ${\delta}E/E$, where
$\delta E \equiv E(t)-E(0)$, as a function of time, and the third column 
exhibits the Fourier transform of the resulting time series
The fourth column exhibits the Lyapunov exponent ${\chi}$, computed for times
${\le}{\;}t$, and the final column exhibits the power spectrum  
$|x({\Omega})|$.

The top row corresponds to a regular Population I orbit.
The second row corresponds to an orbit which remains Population II for a
time $t=4096$. The third row exhibits an orbit which transforms
from Population II to Population III near $t=2048$. If the orbital 
integration for this orbit had been terminated at an earlier time, say 
$t=2048$, the points at large $x$ observed in the leftmost panel of the 
second row would be absent and the power spectrum $|x({\Omega})|$ would have 
substantially less power at very low frequencies. The fourth row exhibits a
typical Population III orbit.
The fact that, for the fourth orbit, ${\chi}$ decreases 
systematically at late times and that most of the power in $x$ is at very
low frequencies manifests the fact that the orbit has been
ejected to very large radii, where the dynamical time is very long and 
typical Lyapunov exponents for chaotic orbits are very small. 

Chaotic and regular initial conditions appear to occupy distinctly different
phase space regions. However, the two different types of chaotic initial
conditions are not so separated: the experiments performed hitherto are 
consistent with the interpretation that, arbitrarily close to every Population
II initial condition, there is a Population III initial condition. 
This suggests strongly that the entropy barriers dividing the two orbit types
may be fractal, exhibiting a structure analogous topologically to cantori in
time-independent Hamiltonian systems (see, {\em e.g.,} Mather 1982).

On the basis of these observations, it appears reasonable to term Population
II orbits as `sticky' chaotic orbits which, like sticky orbits in 
time-independent systems (Contopoulos 1971), are trapped near regular regions;
and, by contrast, to term Population III orbits as `wildly' chaotic orbits,
which exhibit substantial diffusion in energy space.

In any event, as the driving frequency increases, the ensemble eventually 
moves out of the resonance region and all (or almost all) the orbits again 
become regular. If, however, the amplitude of the driving is sufficiently 
large, other, higher frequency resonance regions also exist, in which, once 
again, one observes a coexistence of the three orbit populations. The only
difference here is that the sticky chaotic orbits may be associated with
lower harmonics of the driving frequency, such as ${\omega}/4$ and 
${\omega}/6$. Eventually, however, when the driving frequency
becomes too large, the chaotic orbits disappear for good and the effects of 
the time-dependence `turn off' as the peak frequencies ${\Omega}^{(1)}$
associated with the regular orbits again approach their original values
${\Omega}^{(1)}_{0}.$

\section{THE ROLE OF RESONANCES}

To better understand the role of resonant couplings, it is useful to analyse 
the orbital structure in frequency space by identifying for individual orbits 
the two frequencies in the Fourier spectra, ${\Omega}^{(1)}$ and 
${\Omega}^{(2)}$, which contain the most power. Distinctive patterns relating 
these frequencies to one another and to the driving frequency ${\omega}$ can 
make the resonant nature of the dynamics apparent. In particular, orbits 
locked in a $n:1$ resonance between the natural frequencies and the driving 
frequency lie along the line ${\Omega}^{(2)} = 2 \omega/n - {\Omega}^{(1)}$.
The results for two typical orbit ensembles, one each in the Plummer and
${\gamma}=0$ Dehnen potentials, are provided in Figs.~5 and 6, which exhibit
the peak frequencies generated from $|x({\Omega})|$.

As is illustrated in the bottom row of Fig.~2, for any given energy the peak 
frequencies of
orbits in the unperturbed Plummer and Dehnen potentials,  ${\Omega}^{(1)}_{0}$,
occupy only a small range of values. However, as the driving frequency is
increased from zero, the interplay between the natural frequency and various
harmonics of the driving frequency induces resonant structures. By far the
most dominant is the $2:1$ resonance, which arises when half the driving
frequency ${\omega}$ approaches the range of unperturbed natural frequencies. 
The proximity of ${\omega}/2$ to the natural frequency population causes the
population, confined initially to a thin line (as in panels b of Figs.~5
and 6) to stretch within the confines of the unperturbed range. Eventually,
however, if the amplitude of the driving is sufficiently large, which for 
the potentials here requires $m_{0}{\;}{\ga}{\;}0.2$, there is an onset of
chaos for ${\Omega}^{(1)} - {\omega}/2{\;}{\la}{\;}m_{0}$.

Such a criterion clearly makes sense: If the amplitude of the driving is strong
enough to `wiggle' the frequencies 
to values sufficiently
close to the harmonic of ${\omega}$, frequency overlap occurs and a chaotic
resonance is triggered. This finding thus corroborates the expectation that
chaos arises from an overlap between fundamental frequencies of the unperturbed
system and harmonics of the driving frequency.

If, the Fourier spectra $|x({\Omega})|$ and $|r({\Omega})|$ are overlaid with
a plot of mean values ${\langle}{\chi}{\rangle}$ as a function of ${\omega}$,
one notices that the spacings between peaks are identical. To appreciate the
significance of this fact, one must, however, understand the connection between
 $|x({\Omega})|$ and $|r({\Omega})|$. 
As is evident from the second and fourth rows of Fig.~2, if one denotes by 
${\Omega}_{0}$ the value of ${\Omega}$ associated with the lowest frequency
peaks in 
$|x({\Omega})|$ and $|a_{x}({\Omega})|$, the remaining higher frequency peaks
represent odd harmonics $3 {\Omega}_{0}$, $5 {\Omega}_{0}$, 
. . . . By contrast, the dominant frequency for $|r({\Omega})|$ and
$|a_{r}({\Omega})|$ is 
$2{\Omega}_{0}$ and the higher frequency peaks correspond to multiples 
thereof. This means that $2m:1$ resonances between the driving frequency and 
the natural frequencies 
of motion in the $x$-direction (and, by symmetry, the $y$- and $z$-directions) 
correspond to $m:1$ resonances between ${\omega}$ and the natural radial
frequencies. The spacing between successive peaks in both the $x$- and 
$r$-coordinates is $2 {\Omega}_{0}$.
This is accurate at least to within half the
width of the unperturbed natural frequency range to which
${\Omega}_{0}$ belongs.  

Understanding the onset of chaos, as probed by the mean Lyapunov exponent
${\langle}{\chi}{\rangle}$, as a chaotic resonance between the driving 
frequency
and the range of natural frequencies for the ensemble provides a clear
explanation of why the spacing between peaks in ${\langle}{\chi}{\rangle}$ as
a function of ${\omega}$ are the same as the spacing between peaks in 
$|x({\Omega})|$ and $|r({\Omega})|$ as a function of ${\Omega}$.
Interestingly, however, for lower amplitude pulsations, generally 
$m_{0}{\;}{\la}{\;}0.2$, some peaks in  $|r({\Omega})|$ correspond almost 
exactly to {\em minima} in the amount of chaos although, for larger amplitude,
these minima are overtaken by nearby maxima which emerge as chaos becomes 
more prevalent. If orbits are pulsed with sufficiently small amplitude at a 
frequency that is a harmonic of its natural frequency, the pulsation will 
make them {\em more stable} (in the sense that ${\chi}(t)$ converges faster 
towards zero than for unperturbed orbits), whereas increasing the amplitude 
of pulsations eventually renders them {\em chaotic}.

The resonance structure of the Plummer potential and all the Dehnen potentials,
both cuspy and cuspless, are qualitatively very similar. Resonances occur
whenever the driving frequency ${\omega}$ is close to a harmonic of the
unperturbed natural frequency range, the locations of which vary from model
to model. Besides the dominant $2:1$ resonance, there are other resonances
which cause a significant alignment of orbits along the resonance lines in
frequency space, even when they do not trigger a large measure of chaotic
orbits. For example, as is evident from Figs.~5 m-o and Figs.~6 k-p, both 
the $4:1$ and $6:1$ resonances are clearly present. Their importance and
effect relative to the dominant $2:1$ resonance is commensurate to the relative
power of those frequencies in  $|r({\Omega})|$, which is shown in Figs.~5 a and
6 a.

\section{PHYSICAL IMPLICATIONS}
The analysis described hitherto has focused on distinctions between regular and
chaotic orbits and on determining which resonances are responsible for 
regulating the properties of different orbits.
Attention now focuses on statistical properties of orbit ensembles, and
how these properties depend on the amplitude and the frequency of the 
driving. 

Macroscopically, perhaps the most obvious measure of the efficacy
of the driving is the degree to which it occasions large systematic changes
in energy. In particular, one can ask when the driving causes appreciable 
numbers of orbits to be ejected to infinity, acquiring positive energy. The 
answer here is that, even if the driving is relatively low amplitude, 
for example $m_{0}{\;}{\sim}{\;}0.2$ or less, a sizeable fraction 
of the orbits can be ejected if the driving frequency is in the  
appropriate range. This is illustrated in Fig.~7, which focuses on orbit 
ensembles with initial energy $E=0.6E_{min}$ evolved in pulsed Plummer and 
${\gamma}=0$ Dehnen potentials with amplitude $m_{0}=0.2$ and variable 
driving frequency.

In each case, two different sets of initial conditions were evolved, namely
(i) a microcanonical sampling of the constant energy hypersurface (filled
circles) and (ii) initial conditions corresponding to purely radial orbits
which sampled uniformly the energetically accessible values of $r$ (open
circles).
The top panels exhibit as a function of ${\omega}$ the fraction $f_{e}$ of 
the orbits which have acquired positive energy by $t=4096$, the middle panels 
the mean change in energy at $t=4096$, and the bottom panels mean finite
time Lyapunov exponents. 

It is evident that there is a well-defined `window' of frequencies for which
ejection can occur. For both the Plummer and Dehnen
models, as ${\omega}$ is increased from very small values there is an abrupt
onset of ejections, with $f_{e}$ increasing from zero to its maximum value
over a very short frequency interval. The `closing' of the window at higher
frequencies tends to be more gradual. 

Overall, the frequencies which yield the largest fraction of ejected orbits
and the largest changes in energy tend to correlate with those frequencies 
for which the mean ${\langle}{\chi}{\rangle}$ is especially large. However,
this correlation is not one-to-one. For example, for a microcanonical sampling
evolved in the ${\gamma}=0$ Dehnen potential the largest energy shift occurs 
for ${\omega}{\;}{\sim}{\;}0.8$ but ${\langle}{\chi}{\rangle}$ peaks near 
${\omega}{\;}{\sim}{\;}2.5$. At least in part, such discrepancies arise as
a sampling effect. 

For those frequencies for which driving has the largest effect, 
many of the orbits have been ejected to large radii already at times 
$t{\;}{\ll}{\;}4096$ -- this is, for example, the case in the Plummer 
potential 
for ${\omega}{\;}{\sim}{\;}1$. However, the value of the largest Lyapunov 
exponent scales (at least roughly) inversely with the orbital time scale 
$t_{D}$, so that the short time ${\chi}$ for orbits at large $r$ tends to be 
very small. 
For frequencies where large numbers of orbits have been ejected at early
times, the values of ${\langle}{\chi}{\rangle}$ exhibited in panels (c) and 
(f) reflects an average over a short interval where orbits at relatively 
small $r$ have large ${\chi}$ and a (often much) longer interval where orbits 
at larger $r$ have much smaller ${\chi}$.

The final obvious point evidenced in Fig.~7 is that, in the resonance region, 
as probed both by the degree of chaos and by the typical energy shifts 
experienced, {\em radial orbits tend to be more impacted by the driving than 
`generic' orbits with the same energy}. As is evident from Figs.~7 (b) and (d),
outside the resonant region the microcanonical and radial ensembles behave in 
a fashion which is virtually identical. Inside this region, however, the 
radial orbits are ejected more frequently, experience larger energy shifts,
and have larger finite time Lyapunov exponents. Only near the low frequency
boundary of the resonance region are  the effects of the driving comparable
for radial and generic ensembles. 

That radial orbits are impacted more strongly reflects the fact that they
have significantly more power at higher frequencies than nonradial orbits,
which allows for more efficient resonant couplings. An example of this extra
high frequency power is provided in Fig.~8, which exhibits $|a_{x}({\Omega})|$
and $|a_{r}({\Omega})|$ for representative ensembles of unperturbed radial 
orbits for the same potentials and energies that were sampled microcanonically
to generate Fig.~2.

As evidenced, for example, by Figs.~7 (c) and (f), for comparatively low 
amplitudes $m_{0}$ the degree of chaos, as probed by the mean 
${\langle}{\chi}{\rangle}$, can exhibit a relatively complex dependence on
${\omega}$, although there {\em is} a relatively smooth overall
envelope for ${\langle}{\chi}{\rangle}({\omega})$. However,
as illustrated in Figs.~1 (a) and (b), the short-scale structure tends to 
`smooth out' as $m_{0}$ is increased, this presumably correlating with the
fact that the resonances broadened with increasing amplitude.

By contrast, as illustrated in Fig.~9, for fixed ${\omega}$, the mean 
${\langle}{\chi}{\rangle}$ exhibits a comparatively smooth dependence on 
$m_{0}$. Once $m_{0}$ has become sufficiently large to elicit a significant
response,  ${\langle}{\chi}{\rangle}$ becomes a smooth, monotonically 
increasing  function of $m_{0}$ until the amplitude becomes so large that 
many orbits are ejected to very large radii, where ${\chi}$ become very small. 
Significant also is the fact that, for weak driving, orbits tend oftentimes to 
lose energy and move inwards, whereas, for stronger driving, 
$m_{0}{\;}{\ga}{\;}0.2$, orbits tend generically to gain energy and move 
towards larger radii.

\par\noindent 
\section{CONCLUSIONS AND EXTENSIONS}
The analysis described here focused on the particularly simple case in
which spherically symmetric, integrable potentials are subjected to strictly
periodic perturbations of an especially simple form; and, for this reason,
many of the detailed findings are no doubt nongeneric. The assumption of 
spherical symmetry is useful in that it limits the number of resonances that 
could be triggered, and the fact that the unperturbed potentials are integrable
implies that all the chaos that arises must be be associated with the 
time-dependence. That the perturbation reflects a single large scale 
perturbation  makes the relevant resonances especially easy to identify. And, 
most especially, the seemingly absolute distinction between regular and chaotic
orbits no doubt reflects the strictly periodic form of the time-dependence. 
One knows from other examples (see, {\em e.g.,} Kandrup \& Drury 1998) that 
the introduction of a time-dependent potential that exhibits systematic 
secular changes can occasion transitions between regularity and chaos.

However, several of the conclusions derived from these models are likely
to be robust. Most obvious is the possibility of distinguishing between 
orbits which, over finite time intervals, exhibit exponential sensitivity and 
hence deserve the appellation chaotic, and other, presumably regular, orbits 
that do not. Moreover, there is the fact that, as one might have expected,
orbital structure and the onset of chaos can be associated with specific
resonances, the identities of which could no doubt change in time were one 
to allow for a more complex time-dependence.

It is also apparent that chaotic orbits in a time-dependent potential need 
not exhibit large scale energy diffusion, at least over finite time intervals. 
It may be that, in generic potentials, distinctions between `sticky' and 
`wildly' chaotic orbits are less clear than is the case for the models 
considered here, especially since `sticky' orbits appear tied to specific
subharmonics of the driving frequency, which will vary for a generic 
time-dependence. However, it {\em is} evident that `chaotic' need not 
necessarily imply large scale systematic changes in energy. 
In time-dependent systems exhibiting large amounts of chaos, 
mixing in energies can proceed far less efficiently than mixing in other phase 
space variables.

Relaxing the underlying assumptions entering into this paper could in some
cases make it difficult, if not impossible, to determine precisely the 
resonance structure of orbits. However, there {\em are} a variety of 
interesting issues which can, and should, be addressed. Most obvious, perhaps, 
is the question of what would happen if the unperturbed integrable models 
considered here were replaced by nonintegrable models which admit large 
measures of chaotic orbits even in the absence of a time-dependence. One 
might, for example, expect that such systems would be even more susceptible 
to resonant couplings since the Fourier spectra associated with already 
chaotic orbits are broad band, with power spread over a continuous range of 
frequencies (Kandrup, Abernathy \& Bradley 1995). However, it might
well prove that, in some cases, adding a time-dependence actually converts
chaotic orbits into regular.

It would also be of some interest to determine the sorts of resonances that 
are triggered when a system, integrable or otherwise, is subjected to other 
sorts of periodic time-dependences. For example, an analysis of galactic 
potentials subjected to a (near-)periodic time-dependence associated with a 
supermassive black hole binary shows that the orbital frequencies of the 
binary  which trigger large changes in energy and large 
measures of chaotic orbits need not coincide with the frequencies for which
a global time-dependence $V({\bf r},t)=V_{0}({\bf r})(1+m_{0}\sin {\omega}t)$
triggers large amounts of chaos (Kandrup, Sideris, Terzi{\'c} \& Bohn 2003).

And similarly, as a prolegomena to realistic self-consistent computations,
it would seem important to investigate the effects of large scale variations
in the amplitude and frequency of the time-dependence, for example by allowing
the frequency to drift and/or allowing the amplitude to decrease. Some limited
work on this issue was described in Kandrup, Vass \& Sideris (2003), which
considered bulk oscillations that eventually damp away and random variations
in frequency modeled as colored noise. However, that work was limited
in the sense that attention focused completely on bulk statistical properties
of orbit ensembles, without any consideration of orbital structure. 

From a more formal perspective, it would seem interesting to pursue possible
analogies between phase space topologies in two-degree-of-freedom, 
time-independent  Hamiltonian systems and $1{1\over 2}$-degree-of-freedom
systems (Lichtenberg \& Lieberman 1992) associated with time-dependent, 
spherically symmetric potentials,
for example by reformulating the analysis described in this paper in the
context of an `enlarged phase space' that includes time and energy as 
additional coordinates. It is well known that, in two-degree-of-freedom
time-independent systems, topological considerations preclude the existence
of an Arnold web (Arnold 1964). However one expects generically that the 
chaotic phase space regions near regular islands will be permeated by cantori. 
Analogous considerations would suggest that a  $1{1\over 2}$-degree-of-freedom
system cannot admit an Arnold web, but that some analogue of cantori could 
exist. 

But what might all this mean for real galaxies? Time-dependent potentials 
arise naturally in a variety of astronomical contexts, including long-lived
(quasi-)normal modes, triggered, for example, by close encounters between 
galaxies ({\em e.g.,} Vesperini \& Weinberg 2000) and, most obviously, in 
the violent relaxation of a galaxy involved in a strong encounter.

The normal-mode setting, which involves relative low amplitudes, is more
directly related to the computations described in this paper. Here it would
appear that, as a result of the oscillations, a significant number of orbits
could become chaotic, some exhibiting system drifts in energy and others 
`locked' to a harmonic of the driving frequency. All the chaotic orbits can
participate in chaotic mixing on a constant energy hypersurface. More
importantly, however, the `wildly' chaotic orbits can also mix in energy
space, a process which could trigger a relatively slow (time scale 
${\gg}\,t_{D}$) change in the density distribution. Presuming, as would seem
likely ({\em e.g.,} Kandrup, Sideris, Terzi{\'c} \& Bohn 2003) that this effect
persists in strongly nonspherical systems, this effect could
help drive an asymmetrically-shaped galaxy towards a more axisymmetric shape,
a possibility envisioned by Merritt (1999), or, perhaps, towards some other 
shape that minimizes the amount of chaos (Kandrup \& Siopis 2003). Moreover, 
to the extent that the oscillations persist for a long time, many of the 
`sticky' orbits could become unstuck and, as such, participate in this mixing.

As evidenced in Section IV, even relatively low amplitude driving can also
result in an appreciable number of stars being displaced to very large radii,
and, in some cases, being expelled from the galaxy, presumably in a 
comparatively incoherent fashion. This suggests that, even ignoring direct 
galaxy-galaxy collisions, lower-amplitude galactic `harrassment' could lead 
to the presence of a diffuse population of stars and/or gas in intergalactic 
space.

Violent relaxation differs in that
the amplitude of the time-dependence is substantially higher (at least
initially), that one would expect larger variations in the oscillation 
frequency over time scales ${\sim}{\;}t_{D}$ and, of course, that the
time-dependence should damp systematically on a time scale not much longer
than $t_{D}$. Larger amplitude typically implies broader resonances and,
hence, larger numbers of chaotic orbits with larger finite time Lyapunov
exponent. Indeed, given plausible assumptions it appears likely that all
-- or almost all -- the orbits in some parts of phase space will be chaotic
and, as such, available to participate in chaotic phase mixing (Kandrup, Vass
\& Sideris 2003). Moreover, allowing for a changing oscillation frequency 
need not decrease the abundance of chaos. Indeed, in some cases allowing 
for variations in frequency modeled as coloured noise with a nonzero
autocorrelation time actually increases the relative fraction of orbits 
that are chaotic (Kandrup, Vass \& Sideris 2003). 

The only question is 
whether a sufficiently large fraction of the chaotic orbits will exhibit
secular variations in energy, thus facilitating mixing in energy space. The
answer to this question must of course be addressed in the context of honest
self-consistent simulations. However, it would seem reasonable to expect
that variations in the frequency and the amplitude of the driving might
make the frequency `locking' of sticky chaotic orbits more difficult to
maintain, so that most of the chaotic orbits will exhibit the drifts in
energy required to drive violent relaxation.
\section*{Acknowledgments}
\par\noindent
This research was supported in part by NSF AST-0070809 and AST-0307351.  We 
would like to 
thank the Florida State University School of Computational Science and 
Information Technology for granting access to their supercomputer facilities.

\vfill\eject
\begin{figure}
\centering
\centerline{
    \epsfxsize=8cm
    \epsffile{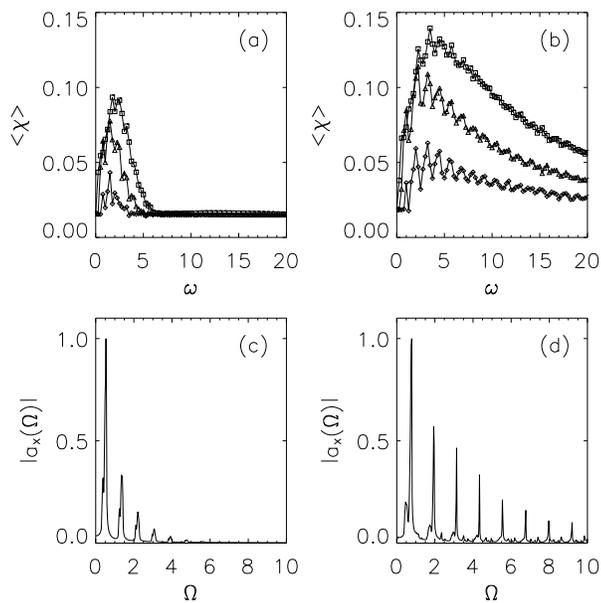}
      }
    \begin{minipage}{10cm}
    \end{minipage}
    \vskip -0.3in\hskip -0.0in
\caption{
(a) Mean finite time Lyapunov exponent ${\langle}{\chi}{\rangle}$ for a 
microcanonical sampling of 1600 orbits with initial energy $E=-0.5$ evolved 
in a Plummer potential subjected
to periodic driving with variable frequency ${\omega}$ with amplitudes 
(from bottom to top) $m_{0}=0.125$, $0.25$, and $0.5$. (b) The same for 
an ensemble evolved in the ${\gamma}=1$ Dehnen potential.
(c) The power spectrum of the acceleration $|a_{x}({\Omega})|$ for 
the same orbits in the Plummer potential, evolved in the absence of driving.
(d) The same for orbits in the ${\gamma}=1$ Dehnen potential with 
$E=-0.5E$. Each orbit was evolved for a time $t=512$.
}
\label{landfig}
\end{figure}

\onecolumn
\begin{figure}
\centering
\centerline{
    \epsfxsize=12cm
    \epsffile{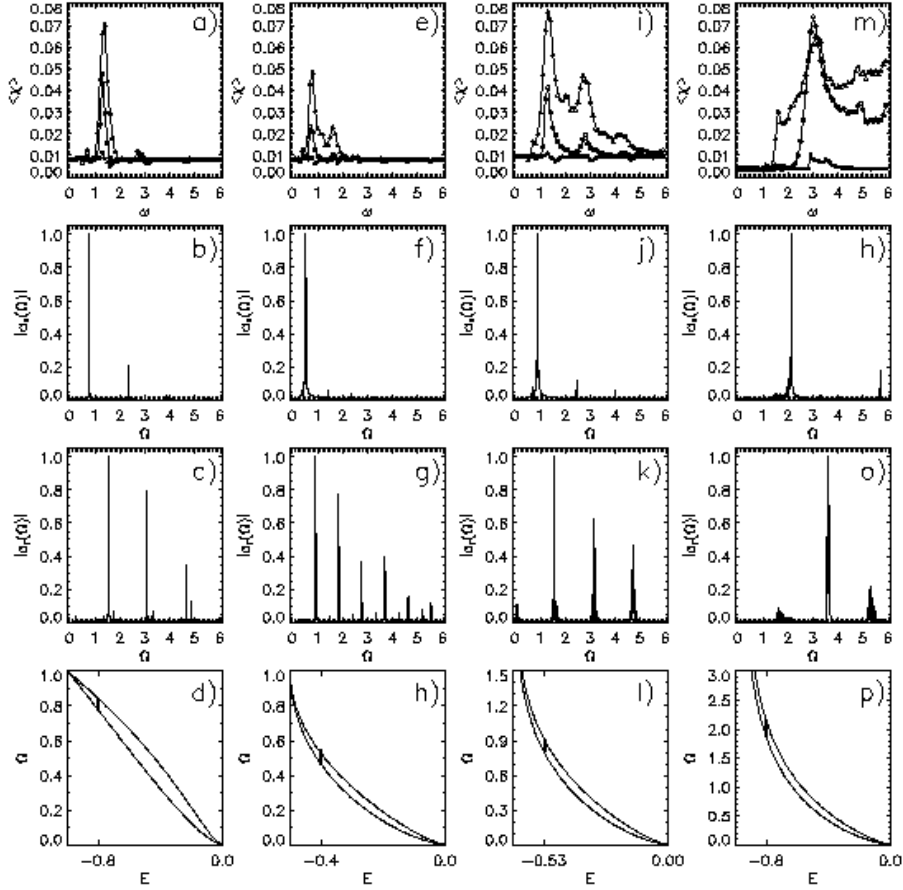}
      }
    \begin{minipage}{10cm}
    \end{minipage}
    \vskip -0.3in\hskip -0.0in
\caption{(a) Mean finite time Lyapunov exponent for a microcanonical sampling 
of 1600 orbits with initial energy $E=0.8E_{min}$ evolved in a Plummer 
potential for a time $t=512$ with variable frequencies ${\omega}$ and 
amplitudes (from bottom to top) $m_0=0.1$ (full circles), $m_0=0.2$ (empty 
circles) and $m_0=0.3$ (triangles). 
(b) $|a_{x}({\Omega})|$, the power spectrum of the $x$-component of the 
acceleration for the same orbits in the Plummer potential, evolved for a
time $t=2048$ in the absence of driving.
(c) $|a_{r}({\Omega})|$, the power spectrum for the radial component.
(d) The analytically computed ranges of frequencies for the unperturbed
Plummer sphere (solid lines), along with the peak frequencies 
${\Omega}^{(1)}$
of orbits integrated in a non-pulsating potential for the initial conditions 
used in time-dependent integrations.
(e) - (h) The same for a pulsating Dehnen potential with ${\gamma}=0$ and
  $E=0.8 E_{min}$.
(i) - (l) For a pulsating Dehnen potential with $\gamma=0.5$.
(m) - (p) For a pulsating Dehnen potential with $\gamma=1.0$.
}
\label{landfig}
\end{figure}
\vskip -.5in
\begin{figure}
\centering
\centerline{
    \epsfxsize=12cm
    \epsffile{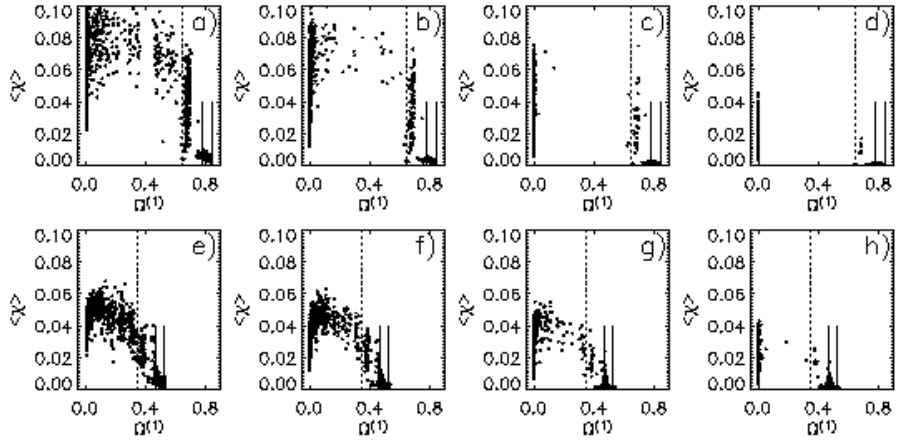}
      }
    \begin{minipage}{10cm}
    \end{minipage}
    \vskip -0.3in\hskip -0.0in
\caption{
Finite time Lyapunov exponents as a function of peak frequency 
${\Omega}^{(1)}$ for Population I (${\Omega}^{(1)}> {\omega}/2$), 
Population II (${\Omega}^{(1)}{\;}{\approx}{\;}{\omega}/2$), and 
Population III
(${\Omega}^{(1)}<{\omega}/2$) orbits, generated from a microcanonical 
sampling
of 1600 initial conditions with $E=0.8E_{min}$, evolved in a pulsating 
Plummer potential with ${\omega}=1.3$ and $m_{0}=0.2$ for times 
(a) $t=1024$, (b) $t=2048$, (c) $t=4096$ and (d) $t=8192$. 
The dashed vertical line corresponds to ${\omega}/2$. The vertical solid
lines show the range of frequencies in the non-pulsating model. The solid
vertical lines delimit the range of unperturbed peak frequencies.
(e) - (h). The same for a pulsating ${\gamma}=0$ Dehnen potential with
${\omega}=0.7$, $m_{0}=0.2$, and $E=0.8E_{min}$.
}
\label{landfig}
\end{figure}

\begin{figure}
\centering
\centerline{
    \epsfxsize=16cm
    \epsffile{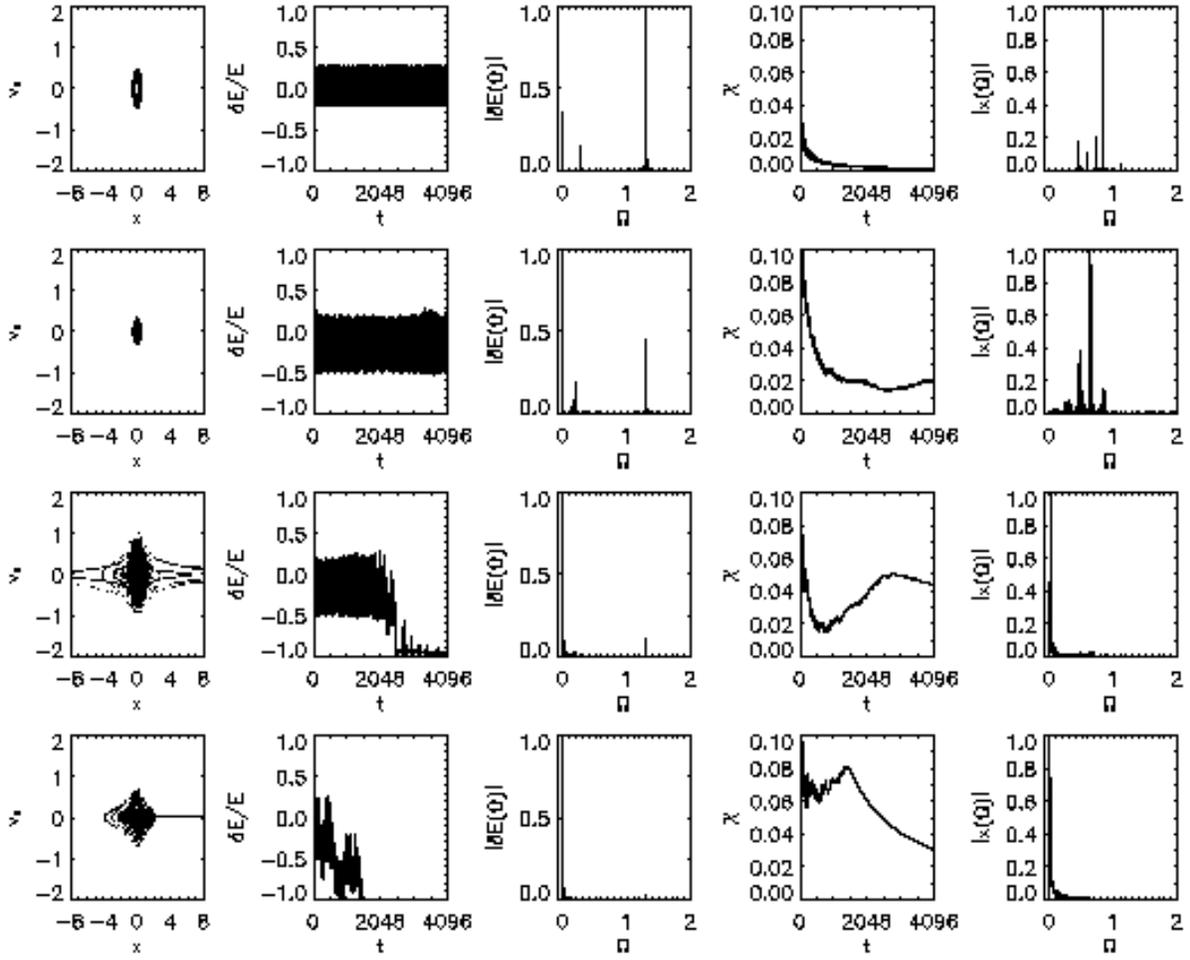}
      }
    \begin{minipage}{10cm}
    \end{minipage}
    \vskip -0.0in\hskip -0.0in
\caption{Top row: The phase space trajectory of a typical Population I radial
orbit evolved in a pulsating Plummer potential with $E=-0.8$, $m_0=0.2$ and 
$\omega=1.3$, along with the relative energy oscillations,
${\delta}E(t)/E$, the power spectrum $|{\delta}E(\omega)|$,
the finite time Lyapunov ${\chi}$ computed for times $T{\;}{\le}{\;}t$,
and the power spectrum, $|x({\Omega})|$.
Second row: The same for a typical weakly chaotic orbit which remains 
Population II for $t<4096$. 
Third row: Another chaotic orbit which transforms from Population II to 
Population III near $t=2048$.
Fourth row: A typical wildly chaotic Population III orbit.
}
\label{landfig}
\end{figure}

\begin{figure}
\centering
\centerline{
    \epsfxsize=14cm
    \epsffile{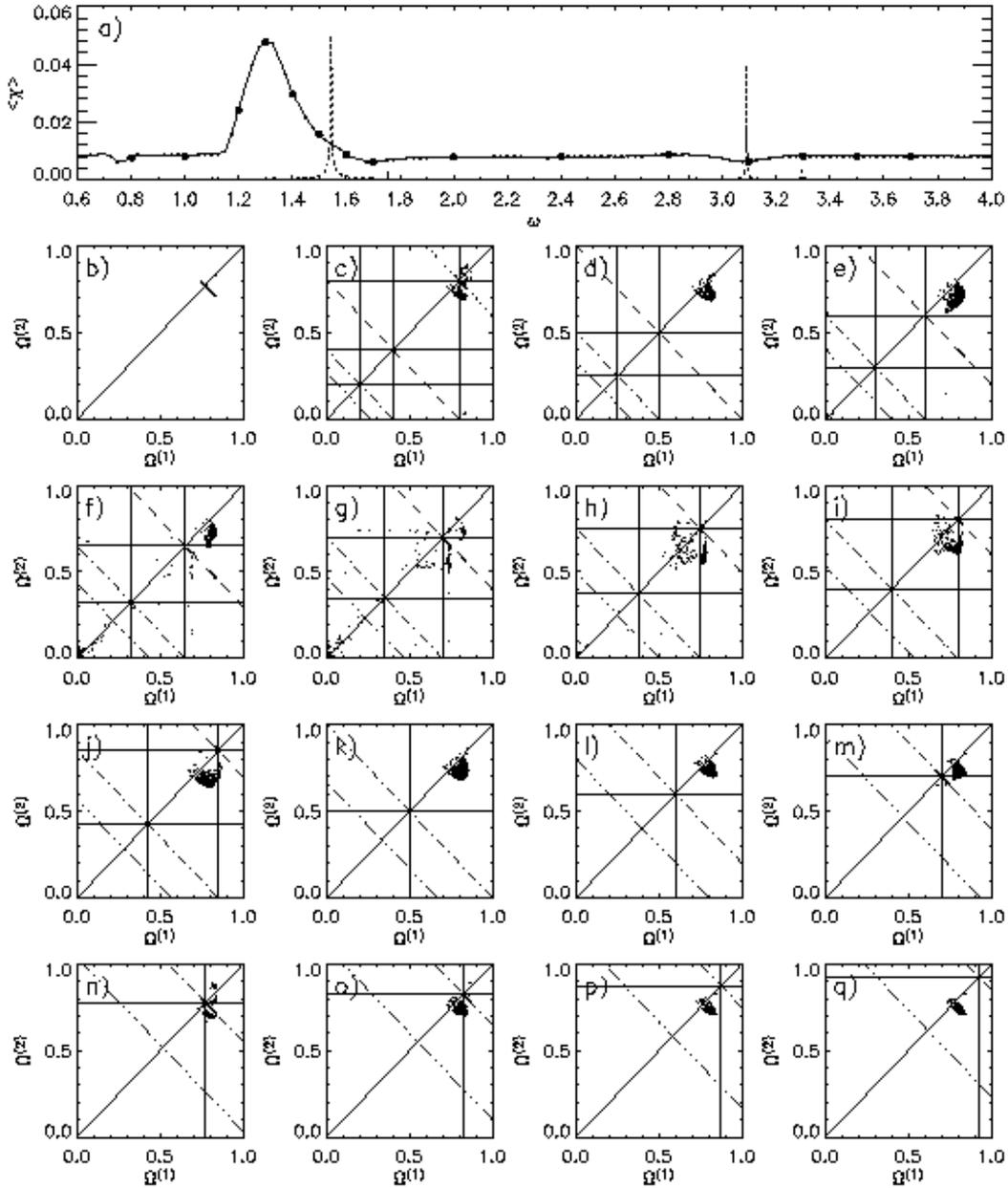}
      }
    \begin{minipage}{10cm}
    \end{minipage}
    \vskip 0.0in\hskip -0.0in
\caption{(a) 
Mean Lyapunov exponents for a microcanonical sampling of 800 orbits with 
initial energy $E=0.8E_{min}$, evolved in a pulsating Plummer potential with 
$m_0=0.2$ for a time $t=2048$. The power spectrum $|r({\Omega})|$ is 
over-plotted with dashed lines. Heavy dots correspond 
to panels (c) - (q), from left to right respectively. (b) A scatter plot in
frequency space, exhibiting the two frequencies $\Omega^{(1)}$ and 
$\Omega^{(2)}$ 
for which, in the absence of any driving, $|x({\Omega})|$ has the most power.
(c) - (q) The same for the pulsating Plummer potential with $\omega = 
0.8, 1.0, 1.2, 1.3, 1.4, 1.5, 1.6, 1.7, 2.0, 2.4, 2.8, 3.1, 3.3, 3.5,$ and 
$3.7$ respectively. The thick diagonal solid line represents 
${\Omega}^{(1)}={\Omega}^{(2)}$.
Vertical and horizontal thin solid lines correspond to the driving frequency 
${\omega}$, ${\omega}/2$, and ${\omega}/4$.  Short-dashed lines represent,
the $1:1$ resonance, long dashes the $2:1$ resonance, dot-dashed lines the 
$4:1$ resonances, and triple-dot-dashed lines the $6:1$ resonance.
}

\label{landfig}
\end{figure}

\begin{figure}
\centering
\centerline{
    \epsfxsize=14cm
    \epsffile{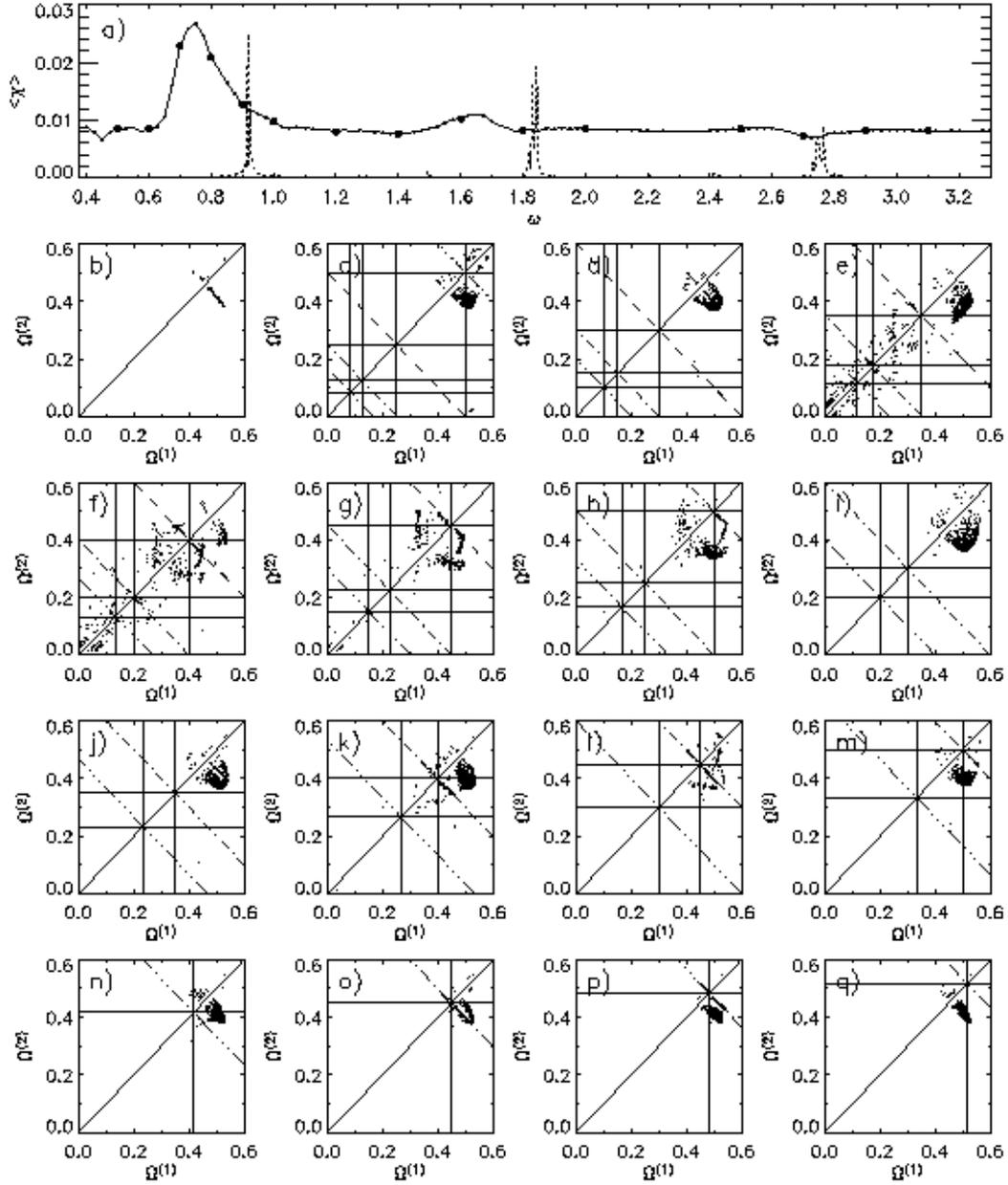}
      }
    \begin{minipage}{10cm}
    \end{minipage}
    \vskip -0.0in\hskip -0.0in
\caption{The same as Fig.~5, albeit for orbits with
initial energy $E=0.8E_{min}$, evolved in a pulsating ${\gamma}=0$ Dehnen 
potential.
}
\label{landfig}
\end{figure}

\twocolumn
\begin{figure}
\centering
\centerline{
    \epsfxsize=8cm
    \epsffile{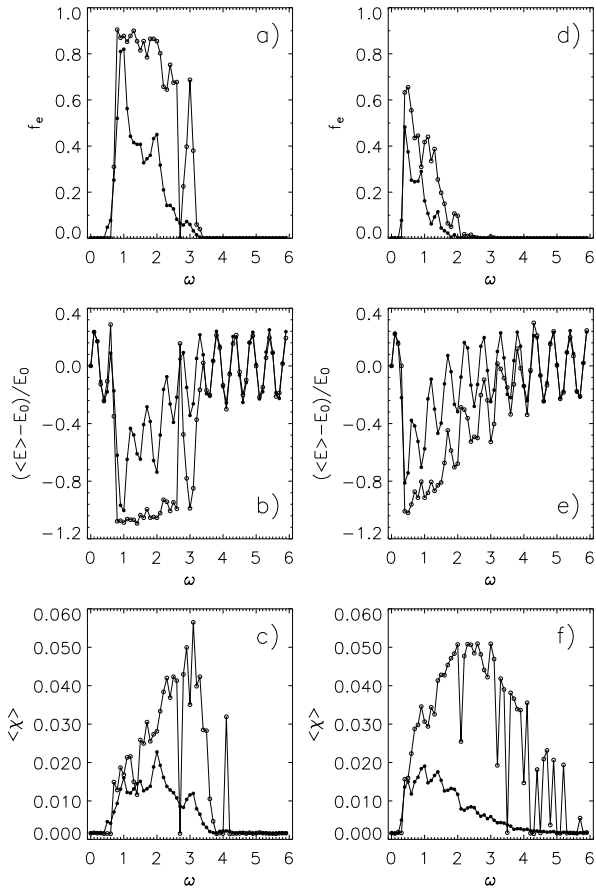}
      }
    \begin{minipage}{10cm}
    \end{minipage}
    \vskip -0.2in\hskip -0.0in
\caption{(a) Fraction of orbits with initial energy $E=0.6E_{min}$, evolved in 
a pulsating Plummer potential with $m_{0}=0.2$ and variable driving frequency 
that have been ejected, {\em i.e.,} acquired positive energy,
by time $t=4096$. Filled circles represent initial conditions sampling a 
microcanonical distribution. Open circles sample a distribution corresponding 
to purely radial orbits. Each point represents $400$ orbits.
(b) Mean relative change in energy for the ensemble after $t=4096$.
(c) Mean finite time Lyapunov exponent for the ensemble.
(d) - (f) The same for an ensemble with $E=0.6E_{min}$, evolved in a 
pulsating ${\gamma}=0$ Dehnen potential.
}
\label{landfig}
\end{figure}

\onecolumn
\begin{figure}
\centering
\centerline{
    \epsfxsize=12cm
    \epsffile{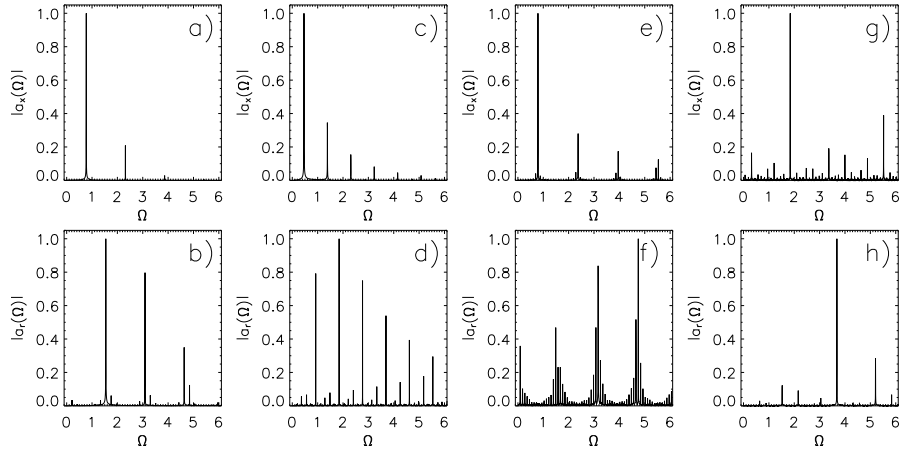}
      }
    \begin{minipage}{10cm}
    \end{minipage}
    \vskip -0.3in\hskip -0.0in
\caption{(a) $|a_{x}({\Omega})|$, the power spectrum of the $x$-component of 
the acceleration for a representative ensemble of radial 
orbits with energy $E=0.8E_{min}$ evolved in the Plummer potential.
absence of driving.
(b) $|a_{r}({\Omega})|$, the power spectrum for the same orbits.
(c) - (d) The same for a pulsating Dehnen potential with ${\gamma}=0$ and
  $E=0.8 E_{min}$.
(e) - (f) For a pulsating Dehnen potential with $\gamma=0.5$.
(g) - (h) For a pulsating Dehnen potential with $\gamma=1.0$.
}
\label{landfig}
\end{figure}
\twocolumn
\begin{figure}
\centering
\centerline{
    \epsfxsize=8cm
    \epsffile{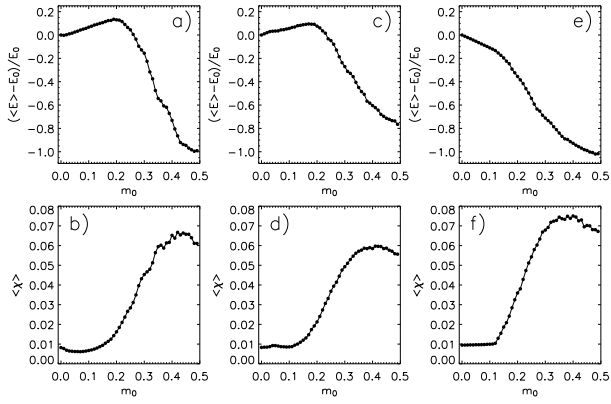}
      }
    \begin{minipage}{10cm}
    \end{minipage}
    \vskip -0.in\hskip -0.0in
\caption{(a) Mean fractional energy shift for a microcanonical sampling of 
1600 
orbits with initial
energy $E=0.8E_{min}$ evolved in a pulsating Plummer potential with frequency
${\omega}=1.5$ and variable amplitude $m_{0}$ for a time $t=4096$. 
(b) The mean finite time Lyapunov exponents computed for the same orbits.
(c) and (d) The same for a pulsating ${\gamma}=0.0$ Dehnen potential with
$E=0.8E_{min}$ and ${\omega}=0.8$.
(e) and (f) The same for a pulsating ${\gamma}=0.5$ Dehnen potential with
$E=0.8E_{min}$ and ${\omega}=1.2$.
}
\label{landfig}
\end{figure}

\end{document}